\documentclass[letterpaper,prd,showpacs,showkeys,10pt,twoside,byrevtex,nofootinbib]{revtex4}

\usepackage[latin1]{inputenc}
\usepackage[english]{babel}
\usepackage{amsmath}
\usepackage{graphics}
\usepackage{hyperref}

\newcommand{\beq}{\begin{equation}}
\newcommand{\eeq}{\end{equation}}
\newcommand{\beqq}{\begin{equation*}}
\newcommand{\eeqq}{\end{equation*}}
\newcommand{\barr}{\begin{eqnarray}}
\newcommand{\earr}{\end{eqnarray}}
\newcommand{\bea}{\begin{eqnarray*}}
\newcommand{\eea}{\end{eqnarray*}}

\newcommand{\dphi}{\delta \phi}
\newcommand{\mH}{\mathcal{H}}
\newcommand{\mP}{\mathcal{P}}

\newcommand{\nk}{\textbf{k}}
\newcommand{\x}{\textbf{x}}

\newcommand{\vac}{|0 \rangle}

\newcommand{\yk}{\hat{y}_k}
\newcommand{\pk}{\hat{\pi}_k}
\newcommand{\ak}{\hat{a}_k}
\newcommand{\aq}{\hat{a}_{k'}}

\newcommand{\bra}{\langle}
\newcommand{\ket}{\rangle}

\newcommand{\ovl}{\overline}
\newcommand{\half}{\frac{1}{2}}
\newcommand{\mpl}{M_{pl}}

\begin{document}

\title{ The  slow roll  condition and the  amplitude  of the primordial   spectrum  of cosmic  fluctuations: Contrasts and similarities of
standard  account and the ``collapse  scheme".}
\author{Gabriel \surname{León}}
\email[E-mail:\,]{gabriel.leon@nucleares.unam.mx}
\affiliation{Instituto de Ciencias Nucleares, UNAM, M\'exico D.F. 04510, M\'exico}
\author{Daniel \surname{Sudarsky}}
\email[E-mail:\,]{sudarsky@nucleares.unam.mx}
\affiliation{ Instituto de Ciencias Nucleares, UNAM, M\'exico D.F. 04510, M\'exico e \\
 Instituto de Astronomía y Física del Espacio, Casilla de Correos  67   Sucursal 28, 1124 Buenos Aires, Argentina.}


\begin{abstract}
The inflationary paradigm enjoys a very  wide acceptance in the cosmological community,  due  in large  part  to  the fact that it is said to ``naturally account" for a nearly scale independent power primordial  spectrum of  fluctuations  which is in very good agreement with the observations. The  expected overall   scale of the  fluctuations  in  most models,  turns out to  be too large,  because it  is  inversely proportional to the  slow roll parameter,  which is  expected to  be   very  small.  This fact requires  the  fine tuning of the  inflaton potential.
In  series  of recent  works it has been  argued  that the   success of the inflationary picture   is not  fully justified in terms of the   rules of quantum theory  as applied to  the cosmological setting  and  that  an extra element,  something akin to a self induced  collapse of the wave function is  required. There,   it  was  suggested that
the   incorporation of   such collapse  in the treatment   might      avoid the need  for  fine tuning of the  potential  that  afflicts   most  inflationary models. In this article we will discuss in detail the  manner  in  which one obtains the  estimation of the  magnitude  of the  perturbations in the new scheme and that of the standard accounts, comparing one of the most popular  among the later and that corresponding to  the new proposal.  We  will see that  the  proposal  includes a collapse  scheme that bypasses  the problem,  but  we  will  see that the price  seems  to a be a teleological, and thus  unphysical,   fine tuning of the  characteristics of   collapse.
\end{abstract}

\maketitle

\section{Introduction}

The  status of the standard inflationary scenario among cosmologists has been dramatically  enhanced  by recent advances in the observations, such as Supernova Surveys \cite{snova}, the studies of large scale structure \cite{lss} and  those arising from the Wilkinson Microwave Anisotropy Probe (WMAP)\cite{wmap}. The model was initially proposed by A. Guth in 1981 \cite{guth} in order to deal with the shortcomings of the standard Big Bang cosmology: namely the Flatness Problem, the Horizon Problem and the Unwanted Relics Problem. The essential idea is that if the Universe undergoes an early era of accelerated expansion (lasting at least some 80 e-folds or so), it would come out of this period as an essentially flat,  homogeneous  and isotropic space-time with an extreme dilution of all unwanted relics. A more dramatic consequence
 \cite{guth2,muk}
 is that when considering  the quantum aspect of  the  scalar field  driving the inflation (the  inflaton field),  which  is   assumed  to  be in the  vacuum state as a result of the same  exponential expansion,  one finds that  it contains   ``fluctuations" with the   appropriate scale-free -Harrizon-Zeldovich  spectrum. These vacuum fluctuations,  are  thus  considered responsible for all the structure we observe in the actual universe, and in particular, the observed Cosmic Microwave Background (CMB) anisotropies.
The shape of the observed spectrum
  turns out to be in excellent agreement -when adjusted to take into account well established physics of plasma oscillations and related phenomena- with a very  flat  primordial spectrum  fluctuations. The  shape of the spectrum that results from the   general picture emerging  from  these theoretical considerations, is controlled   by  the so called  ``slow  roll parameter" $ \epsilon$  (to  be  precisely defined  below),  in  such a way that the  spectrum becomes  flatter   as  that parameter is decreased,  on the other hand the overall scale is  inversely proportional to the
 parameter   $\epsilon$  indicating that  it can not be made too small.

The details of the model, such as the form of the potential, the number of fields, etc, continues to be a subject of extensive research, while much less attention has been devoted to the question on how exactly does the universe transit from a homogeneous and isotropic stage to one where the quantum uncertainties, which are after all still homogeneous and isotropic, become actual inhomogeneities.

In  this  regard, a  recent series of works \cite{sud,sud2}  have argued  that the standard inflationary paradigm has an important shortcoming. The point is that the scheme  by itself does not provide a fully satisfactory explanation for  emergence of the seeds  of  structure, as there is nothing in the theory that could  account for the transmutation of  the initial\footnote{  By  this  we  refer  here to the state of the universe   after  just   a few  e-folds of  inflationary  expansion, and not to the  state that precedes  inflationary regime,  presumably   emerging from the Planck   regime, and   which is  expected to   involve all  sorts of inhomogeneities and  defects,  which inflation is  assume  to  erase.} homogeneous and isotropic  situation (as described  by the background and the quantum state  of the fields corresponding to the early universe) into an in-homogeneous and an-isotropic situation corresponding to the latter state  representing   the actual universe we observe.  \emph{Quantum decoherence} has been a constant reference  in the attempts to deal with this issue (also referred as the ``quantum to classical transition"  \cite{kiefer}),  but a careful analysis  of the resulting accounts  indicates that such  approaches  fail to  provide a true  resolution  \cite{shortcomings}.

We  will not  dwell here on these conceptual issues  and   will only illustrate the problem  by  quoting from page 476 of Cosmology by S. Weinberg \cite{weinberg}:

\begin{quotation}
\emph{These are quantum averages, not averages over an ensemble of classical field configurations. ... Just as in the measurement of a spin in the laboratory, some sort of decoherence must set in; the field configurations must become locked into one of an ensemble of classical configurations, ... It is not apparent just how this happens, ...}
\end{quotation}

Regarding a  widespread  belief that  one can  address the issue   invoking decoherence,
we  will again   limit  ourselves, to just  quote from pages 348, 349 of Physical Foundations of Cosmology by  V. Mukhanov \cite{mukbook}:

\begin{quotation}
\emph{How do quantum fluctuations become classical? ... Decoherence is a necessary condition for the emergence of classical inhomogeneities and can
easily be justified for amplified cosmological perturbations. However, decoherence is not sufficient ... It can be shown that as a result of unitary evolution we obtain a state which
 is a superposition of many macroscopically different states, each corresponding to a particular realization of galaxy distribution. Many of these realizations have the same statistical properties. ... Therefore, to pick an observed macroscopic state from the superposition we have to appeal either to Bohr's reduction postulate or to Everett's many-worlds interpretation of quantum mechanics. The first possibility does not look convincing in the cosmological context....}
\end{quotation}

%
Thus it is clear that the standard description,  within  inflationary  cosmology,  of  the origin of  the seeds of cosmic  structure, is  far from providing a fully  satisfactory account
and faces   important conceptual  shortcomings.

In order to overcome such shortcoming  of the standard inflationary model, the authors in \cite{sud} introduce a new ingredient to the inflationary paradigm: the {\it  self induced collapse hypothesis}: a  phenomenological model incorporating the description of the  effects of a  dynamical collapse of the wave function of the  inflaton on the subsequent  cosmological evolution. The  idea  is inspired  by R. Penrose's  arguments in the  sense  that  the unification of  quantum  theory and the theory of gravitation  would  likely involve  modifications in both theories,   rather  than only the latter  as is  more frequently assumed.   Moreover the  idea  is  that the resulting modifications of the former should involve something akin to  a self-induced  collapse of the  wave-function   occurring when the matter fields  are in  a  quantum superposition that would  lead  to  corresponding space-time  geometries  which   are ``too different  among themselves". 
This  sort of self induced  collapse  would  in fact  be occurring   in   rather  common  situations, and would ultimately resolve the long standing  problem  known as the ``measurement problem" in quantum mechanics
 \cite{penrose, penrose2}.   We  will  not   further  discuss  these motivations here. In our treatment, which can be   seen as  an attempt to realize  these  ideas in the cosmological setting,  as a means  to resolve the   above mentioned shortcoming,   the actual formalism  must   be    considered  as  an effective  description of the  fundamentally quantum gravitational   mechanism,
 which, in our situation leads to the transition  from the  symmetric  vacuum state to the asymmetrical (the symmetry being homogeneity and isotropy) latter state. At this stage   the analysis  should  be seen as  a  purely phenomenological  scheme, in the sense that it does not attempt to explain such collapse  in terms of  some specific  new physical mechanism, but merely gives a rather general parametrization of such a transition.  We  will refer  to this phenomenological model  as the  \emph{collapse scheme}. We will not  further recapitulate the motivations and discussion of the original proposal and instead refer the reader to the above mentioned works.

The issue that concern us in this paper is
 the fact that in order to obtain a suitable fluctuation spectrum, the standard inflationary scenarios  require  the  slow  roll parameter,  to be on the one  hand   small enough to give a flat  spectrum,  and on the  other hand    has to  be large enough
to ensure that the fluctuations  are as ``small" as observed.  Sometimes  this  is   presented  as indicating that  the  scale of the  inflaton potential has to  be carefully fine tuned. In  some early works  on the  collapse  scheme   it was argued \cite{sud,sud2} that the new approach  seemed to   offered  a  possibility to  solve that fine-tuning problem. This  is the subject that will  occupy us here.

 In this  manuscript, we  will  review the  manner  in  which one obtains the  estimation of the  magnitude  of the  perturbations both in the standard approach and  compare it  with the  corresponding analysis  in the collapse scheme.  We will see that, in the standard  paradigm, one  obtains  an expression for the ``power spectrum"  which is proportional to $V/\epsilon$.
  On the other hand, in the collapse scheme, one can find a very particular set of  characteristics for the collapse that  would   indeed avoid the need for that adjustment. That is, the collapse scheme admits a specific  model  for  the collapse  which leads to a fluctuation spectrum with the shape and the amplitude which  move  together in the appropriate  direction as   $\epsilon$ is  decreased:  the flatter the spectrum  the smaller its amplitude.  This  is  in accord  with the  earlier suggestions, however as it will be shown here, this is not a generic feature of the collapse scheme, and in most cases one will end up with an amplitude proportional to $V/\epsilon$. At this  time  we  see no  natural  way by  which   the  mechanism   would   select  the particular   characteristics,   and    the  only  way  of  requiring  such behavior   seems to  involve an  adjustment  of the  parameters of the collapse to  the details of the  reheating,  in  what  can only  be  described as a  teleological   arrangement.

  We  should note  that in the collapse  model   as  developed in \cite{sud},  the scalar and tensor  fluctuations  end  up  having     very  different  spectra, with the former  acquiring a nontrivial  value as a result of the collapse of the wave function of the matter field  and its effect on the so called  Newtonian Potential,  while the  latter end  up  being absent completely (at least at the  first order  of perturbation  theory,  which  is  what  has been  considered so far)  due to the fact that  there  are,  at  that order, no matter sources  generating the tensor  perturbations. The difference  can, in this case, be traced to the fact that  background is  (in the appropriate coordinates) ``time-dependent" but does not  depend  on the spatial  coordinates (i.e.  is  homogeneous and isotropic). For details  on this issue  we  refer the  reader to the cited manuscript.

The article is organized as follows: In Section \ref{mukhanov} we review the standard description of the inflationary scenario, making special emphasis in the steps where aspects connected with the scale of the  resulting  spectrum
 make  their appearance, ending with the calculation of the power spectrum of the metric fluctuations.
  In  section \ref{Earlier}   we  review  the earlier analysis   leading to the suggestion that  the collapse  scheme might  naturally  resolve the problem. In Section \ref{sudarsky} we briefly review the quantum mechanical treatment of the field's fluctuations within the collapse scheme and show how the proposal,  offers a path  that  changes the conclusions,  a  path however that,  as  indicted above, seems at this point  rather unconvincing.
In Section \ref{discusion} we end with a brief  discussion of our conclusions.

Regarding notation we will use signature $(-+++)$ for the metric and Wald's convention for the Riemann tensor.

\section{The standard inflationary scenario} \label{mukhanov}

This section will  briefly review  the  standard inflationary scenario following closely Chapter 8 of  \cite{mukbook}. We elected to  focus on  this reference because it serves as a pedagogical introduction to the subject and   because \cite{muk} is considered as one  of  the standard  references on inflationary quantum perturbations.  We will pay particular attention  to aspects connected to the  estimation of the  overall  scale of the  fluctuation spectrum.

 In the standard inflationary model  the early universe is dominated by a scalar field $\phi$ with a particular  potential $V(\phi)$ called the inflaton. This  potential  acts as a cosmological constant, which   is later ``turned off"  (when the inflaton  reaches the zero of the potential) as a result of the scalar field  dynamics, followed  by a reheating  period   ``bringing back"  the universe to the standard Big Bang cosmological evolutionary path.

 The model  is  characterized  by the action of a scalar field coupled to gravity:

\begin{equation}\label{actioncol}
S[\phi] = \int d^4x \sqrt{-g} \bigg[ \frac{1}{16 \pi G} R[g] -\half \nabla_a \phi \nabla_b \phi g^{ab} - V[\phi] \bigg].
\end{equation}

The analysis  starts  with a  background  space-time that,  as the result of  the exponential  inflationary  expansion, has been  driven to a  homogeneous and isotropic stage, characterized by the  space time  geometry described,  in accordance  with the  general  inflationary  paradigm,  by the  spatially  flat RW cosmology  and  a   background  scalar field:

\beq
ds^2 = a^2(\eta)[-d\eta ^2 + \delta_{ij}dx^i dx^j], \qquad  \phi_0 (\eta).
\eeq

 The    background  metric and  the  field  up to this point  represent an   homogeneous-isotropic classical \emph{background} (or ``expectation value").
Next  one considers the quantum  aspect  of the field $\dphi (\x,\eta)$, representing the  \emph{quantum fluctuations}, and  their  effect of the  space-time geometry $\delta g_{ab}$.

The  energy momentum  tensor involves   both the  inflaton field  and radiation.  The first  dominating the early  inflationary  era   which  ends  in  a reheating  period  after which  the radiation dominates. The  treatment  can be done separately for each regime, but
one  can simplify the treatment  by using in general  the expressions appropriate  for the perfect fluid with energy-momentum tensor $T_{ab}= (\rho+P)U_aU_b + P g_{ab}$ and   when considering the inflaton's    contributions   using  the identification:    $\rho=X + V$, $P= X - V$, where $X \equiv - \frac{1}{2} g^{ab}\partial_a \phi \partial_b\phi$
 and $ U^a =  -g^{ab}\partial_b \phi/\sqrt{2X}$.

 Einstein's equations for the background  are written as    $G_{00}^{(0)}=8\pi G T_{00}^{(0)}=8\pi G a^2 \rho$ and  $G_{ii}^{(0)}=8\pi G T_{ii}^{(0)}=8\pi G a^2 P$  and  yield  Friedmann's equations:

\beq\label{friedmann1}
3\mH^2=8 \pi G a^2 \rho, \qquad   -2\mH' - \mH^2 = 8\pi G a^2 P,
\eeq

where $\mH \equiv a'(\eta)/a(\eta)$; the prime denotes derivative with respect to the conformal time $\eta$; $\rho$ is the overall energy density while  $P$ is  the overall pressure.

Friedmann's equations \eqref{friedmann1}  can be combined to yield a useful expression for $\rho + P$:

\beq\label{friedmann2}
\mH^2 - \mH' = 4\pi G a^2 (\rho + P).
\eeq

 As indicated, in the general setting these densities and pressures  include the contributions of both the inflaton field and of  other forms  of matter and radiation that might be present.


The evolution equation for $\phi_0$ is:

\beq\label{1000}
\phi_0'' + 2 \mH \phi_0' +a^2 \partial_\phi V = 0.
\eeq

The equations above lead to the standard inflationary regime, which written using conformal time, is characterized  by  a scale factor $a(\eta) \approx -1/[H_I(1-\epsilon) \eta]$, with $H_I^2 \approx 8 \pi G V/3$; $\epsilon \equiv 1 - \mH'/\mH^2$ the slow-roll parameter (characterized by $\epsilon \ll 1$ during inflation)  and with the scalar field $\phi_0$ in the slow roll regime so $\phi_0' = -(a^3 / 3a') \partial_\phi V$. This era is supposed to end   while giving rise to a
``reheating period" whereby the universe is repopulated with ordinary matter fields, and then, to a standard hot big bang cosmology leading up to the present cosmological time.

The normalization of the scale factor will be set so $a=1$ at the ``present cosmological time". The inflationary regime would end at $\eta = \eta_r$,  a value  which is negative and very small in absolute terms ($\eta_r \approx -10^{-22}$ Mpc), that is, the conformal time during the inflationary era is in the range $-\infty < \eta < \eta_r$, thus $\eta=0$ is a particular value of the conformal time that does not correspond to the inflationary period, in fact it belongs to the radiation dominated epoch. The scale factor evaluated at the end of the inflationary regime would be denoted as $a_r \equiv a(\eta_r)$.



 Next   one   considers Einstein's equations to first order in the perturbations\footnote{As always when working in perturbation theory one should be worried about issues of linearization  stability. Fortunately the studies of these issues  in flat   cosmological models  involving matter fields \cite{bruna},  indicate that  there would not be a problem. However, in the  case of a closed  cosmological model there might be some troubles and, in the face of  the results of \cite{losic}, the  issue   should be considered  with care. Also the situation seems to become more complex when, in contrast with what is done in this manuscript,  one attempts to  treat the linearized gravitational sector in a quantum level.}. According to the usual  approach, at this  point  it  is argued  that these quantum  fluctuations  will  result in the space-time metric  developing  anisotropies  and inhomogeneities.
It is customary to decompose the metric fluctuations in terms of its scalar, vector, and tensor components. In the case of the Einstein-inflaton system,  we need to concern ourselves  only with  scalar  and tensor perturbations.  In fact the latter
will be ignored by simplicity, therefore we can write  the perturbed metric (in the longitudinal gauge) as:

\beq\label{metric}
ds^2 = a(\eta)^2 [-(1+2\Phi) d\eta^2 + (1-2\Psi) \delta_{ij} dx^idx^j],
\eeq

where $\Phi$ and $\Psi$ are   functions  of  the space-time coordinates $\eta, x^i$, with the former  referred to as the Newtonian potential.

 The equations  $\delta G_0^0 = 8 \pi G \delta T_0^0$ and $\delta G_i^0 = 8 \pi G \delta T_i^0$, are given respectively by

\beq\label{00}
\nabla^2 \Psi -3\mH(\mH\Phi + \Psi') =- 4 \pi G a^2 \delta T_0^0,
\eeq
\beq\label{0i}
\partial_i (\mH \Phi + \Psi') = -4 \pi G a^2 \delta T_i^0.
\eeq

On the other hand, equation $\delta G_j^i = 8 \pi G \delta T_j^i$ is:

\beq\label{ij}
 [\Psi'' + \mH(2\Psi+\Phi)' + (2\mH' + \mH^2)\Phi + \textstyle{\half} \nabla^2 (\Phi - \Psi)] \delta^i_j - \half \partial^i \partial_j (\Phi - \Psi) =  4 \pi G a^2 \delta T_j^i.
\eeq

 For the situations of interest  in this  work,  we  can write  generically $ \delta T_0^0=- \delta \rho$,   $\delta T_i^0 =  (\rho+P) U^0 \delta U_i$    and  $\delta T_j^i= \delta P \delta_j^i$.

 It  is  easy to see that  consideration  of \eqref{ij} for
 the case $i\not=j$,   together with a ppropriate  boundary conditions (more easily seen in the Fourier transformed  version)
  leads to  $\Psi = \Phi$. From now on we will use this result.

For generic hydrodynamical matter, the pressure is a function of both the energy density $\rho$ and the ``entropy per baryon" $S$, and hence:

\beq\label{deltaP}
\delta P = c_s^2 \delta \rho + \tau \delta S,
\eeq

with $c_s^2 \equiv (\partial P / \partial \rho)_S$ the adiabatic speed of sound and $\tau \equiv (\partial P/ \partial S)_\rho$. The expressions \eqref{00}, \eqref{ij}  and \eqref{deltaP} can be combined to yield the following  equation of motion for $\Psi$

\beq\label{psimaster}
 \Psi'' - c_s^2 \nabla^2 \Psi + 3\mathcal{H} (1+c_s^2) \Psi ' + [2\mathcal{H}' + \mathcal{H}^2 (1+3c_s^2)] \Psi = 4\pi G a^2 \tau \delta S.
\eeq

It will be convenient to recast   \eqref{psimaster} in a slightly different form,
  with  the introduction of  the new variable:

\beq\label{u}
u\equiv \frac{\Psi}{4\pi G \sqrt{\rho  + p}}.
\eeq

Using the continuity  equation $\rho'= -3\mH (\rho + P)$  and the definition of $c_s^2$ one finds that $1+c_s^2 = -(\rho' + P')/[3\mH(\rho+P)]$. Substituting this last expression in \eqref{psimaster} and changing the variable $\Psi$ for $u$, the evolution equation for $u$ can be written in the form:

\beq\label{umaster}
u'' - c_s^2 \nabla^2 u - \frac{\theta''}{\theta} u = \mathcal{N},
\eeq

where $\theta \equiv \mH/a^2(\rho + P)^{\half}$ and $\mathcal{N} = a^2 \tau \delta S / (\rho +P)^{\half}$. The motion equation \eqref{umaster}
 or equivalently \eqref{psimaster}  is very useful because it allow us to determine the evolution of perturbations
 in a hydrodynamical universe, and are the main result of section 5 in \cite{muk}.

For the inflationary era, there is an equation  similar to \eqref{psimaster},  which  can be obtained by
 following the corresponding  steps that lead us to the later. That is, from the   general  expression for the energy-momentum tensor for the inflaton.
$T^a_{b} =g^{ac} \partial_c \phi \partial_b \phi + \delta^a_b ( -\half g^{cd} \partial_c \phi \partial_d \phi - V(\phi) )$
one  obtains  the components of the linear perturbations:

\begin{subequations}
\beq
 \delta T^0_0 = a^{-2}[\phi_0'^2 \Phi - \phi_0' \dphi' - \partial_\phi V a^2 \dphi], \qquad \delta T^0_i = \partial_i (-a^{-2} \phi_0' \dphi),
\eeq

\beq
\delta T^i_j= a^{-2}[  \phi_0' \dphi' -\phi_0'^2 \Phi - \partial_\phi V a^2 \dphi]\delta^i_j.
\eeq
\end{subequations}

Thus, Einstein's equations for the  metric  perturbations \eqref{00}, \eqref{0i}, \eqref{ij} take the form:

\beq\label{00inf1}
\nabla^2 \Psi -3\mH(\mH\Phi + \Psi') = 4 \pi G [-\phi_0'^2 \Phi + \phi_0' \dphi' + \partial_\phi V a^2 \dphi],
\eeq

\beq\label{0iinf1}
\partial_i (\mH \Phi + \Psi') = 4 \pi G \partial_i ( \phi_0' \dphi),
\eeq

\barr\label{ijinf1}
 [\Psi'' + \mH(2\Psi+\Phi)' + (2\mH' + \mH^2)\Phi &+& \textstyle{\half} \nabla^2 (\Phi - \Psi)] \delta^i_j - \textstyle{\half} \partial^i \partial_j (\Phi - \Psi) = \nonumber \\
& &  4 \pi G [\phi_0' \dphi' -\phi_0'^2 \Phi  - \partial_\phi V a^2 \dphi]\delta^i_j.
\earr

Subtracting \eqref{00inf1} from \eqref{ijinf1}, using:  $\Psi = \Phi$,  the   equation for the background field \eqref{1000} and \eqref{0iinf1}, one finds  the following equation  of motion for $\Psi$ during the inflationary regime:

\beq\label{psimasterinf1}
\Psi'' - \nabla^2 \Psi + 2 \bigg( \mH - \frac{\phi_0''}{\phi_0'} \bigg) \Psi' + 2 \bigg( \mH' - \frac{\mH \phi_0''}{\phi_0'} \bigg) \Psi = 0.
\eeq

Thus, by  setting

\begin{equation}\label{cs2tds}
c_s^2 = \frac{-1}{3} \bigg( 1 + \frac{2\phi_0''}{\mathcal{H} \phi_0'} \bigg) \qquad \mbox{and} \qquad  \tau \delta S =\frac{(1-c_s^2) \nabla^2 \Psi}{4\pi G a^2},
\end{equation}

in \eqref{psimaster}, one obtains  \eqref{psimasterinf1}. Equivalently, substituting \eqref{cs2tds} in  \eqref{umaster} one obtains

\beq\label{umasterinf}
u'' - \nabla^2 u -\frac{\theta''}{\theta} u = 0.
\eeq

Thus we have an equation for the Newtonian potential,  \eqref{psimaster} (or equivalently \eqref{umaster})  which   is valid  in the inflationary or matter  dominated  regimes,  as long as  in the  case of the former one  uses  the expressions for the ``adiabatic sound speed" and the ``perturbations to the entropy" as given by \eqref{cs2tds}.  One should note however,  for the  case of the  scalar field, the thermodynamical notions  such as  ``equation of state",   ``entropy per baryon"  or  ``sound speed" must  be  interpreted  with care, as  in that case, we are not  really dealing with hydrodynamical matter (see \cite{malik} for a detailed discussion).

We will now proceed to the  treatment of the quantum part of the model, by focusing on   the quantum  ``fluctuations" of the inflaton.   The proceeding discussion will follow the analysis presented in \cite{mukbook}.

Einstein's equations \eqref{00inf1} and \eqref{0iinf1} can be rewritten as:

\beq\label{855}
\bigg( a^2 \frac{\Psi}{\mH} \bigg)' = \frac{4 \pi G a^4(\rho+P)}{\mH^2} \bigg( \mH \frac{\dphi}{\phi_0'}+\Psi \bigg),
\eeq

\beq\label{854}
\nabla^2\Psi = \frac{4\pi G a^2 (\rho + P)}{\mH}  \bigg( \mH \frac{\dphi}{\phi_0'}+\Psi \bigg)'.
\eeq

 Next one  notes that  \eqref{855} and \eqref{854},  can be expressed  in a convenient  way in terms of the variable $u$ (introduced in \eqref{u}) and a  new   variable $v$:

\beq\label{856}
v \equiv a\bigg( \dphi + \frac{\phi_0'}{\mH} \Psi \bigg),
\eeq

whereby  they  take the simple form:

\beq\label{857}
\nabla^2 u = z \bigg( \frac{v}{z} \bigg)', \qquad v=\theta \bigg( \frac{u}{\theta} \bigg)',
\eeq

with $z \equiv \theta^{-1} \equiv a^2(\rho + P)^{1/2}\mH^{-1}$. The next step in the  usual approach is to  write the quantum theory  for  these  field variables.  This is  done writing the second order perturbation  expansion  for  the action
of the gravitational and scalar fields about the background.  Dropping total derivative terms, the result reduces to a simple  expression containing only the variable $v$.

\beq\label{879}
S \equiv \int d\eta d^3x \mathcal{L} = \frac{1}{2}\int d\eta d^3x \bigg( v'^2+v \nabla^2 v+\frac{z''}{z} v^2 \bigg).
\eeq

The details  of these  calculations can be seen  in section 10 of \cite{muk}. This  action leads to the following  simple  equation of motion for $v$,

\beq\label{880}
v''-\nabla^2v-\frac{z''}{z}v  = 0,
\eeq

indicating (see page 341 of \cite{mukbook}) that  the quantization of cosmological perturbations with  the action \eqref{879} is formally
equivalent to the quantization of a ``free scalar field" $v$ with time-dependent ``mass" $m^2 = -z''/z$ in Minkowski space.



 The general solution  for  the operator version of the   equation  of motion \eqref{880},  (as appropriate   when  viewing  the  quantum theory in the Heissenberg picture) can be written as:

\beq\label{886}
\hat{v}(\eta,x) = \frac{1}{\sqrt{2}} \int \frac{d^3k}{(2\pi)^{3/2}} [v_k^\star (\eta)e^{ikx} \hat{a}_k^- + v_k (\eta)e^{-ikx} \hat{a}_k^+],
\eeq

where the temporal mode functions $v_k(\eta)$ satisfy:

\beq\label{885}
v_k''+\omega_k^2(\eta)v_k = 0, \qquad \omega_k^2 \equiv k^2-z''/z.
\eeq

The normalization condition for the mode functions $v_k(\eta)$ are chosen  so that:
$v_k' v_k^\star - v_k v_k^{\star'}  = 2i$,
 which leads to the standard commutation relations for the annihilation and creation operators $\hat{a}_k^{-}$ y $\hat{a}_k^{+}$:
 $[\hat{a}_k^-,\hat{a}_{k'}^-]=[\hat{a}_k^+,\hat{a}_{k'}^+]=0, [\hat{a}_k^-,\hat{a}_{k'}^+]=\delta (k-k')$


The construction is   fully defined   by    the selection  of
  specific  mode  functions specified  at  an ``extremely early" time $(\eta_i)$  by  the initial  conditions:
%
%
%
$v_k(\eta_i) = w_k^{-1/2}$,
$ v_k'(\eta_i)=iw_k^{1/2}$
leading to a   construction  of the quantum theory where  the vacuum state  is the so  called  Bunch-Davies  vacuum \cite{davies}, which is  supposed  to  describe  the state of the quantum field, well  into  the inflationary regime.

Next   one  proceeds to  evaluate the two point correlation function, and then  interprets it  as the power spectrum of the gravitational potential.  In other words,
the analysis is  based on the identification:

\beq\label{correlacion}
\bra 0| \hat{\Psi}(\eta,x) \hat{\Psi} (\eta,y) \vac   \equiv  \ovl{\Psi(\eta,x) \Psi(\eta,y)},
\eeq
 from which one obtains  the ``prediction" for  the primordial spectrum of  cosmic  structure.  The  problems  related  to the  justification of  such identification are  closely connected  with  the conceptual problem we mentioned  in the introduction, and  as indicated  will not be further   discussed  here. We  will instead   focus  on obtaining  the  estimation of the overall   scale of the power spectrum in the standard approach:

From  \eqref{u}, \eqref{857} and \eqref{886} we obtain an expression for the operator $\hat{\Psi}$:

\beq\label{894}
 \hat{\Psi}(\eta,x) = \frac{4\pi G(\rho+P)^{1/2}}{\sqrt{2}}\int \frac{d^3k}{(2 \pi)^{3/2}}[u_k^\star (\eta)e^{ikx} \hat{a}_k^- + u_k (\eta)e^{-ikx} \hat{a}_k^+],
\eeq

where the mode functions $u_k(\eta)$ satisfy:

\beq\label{859}
u_k''+ \bigg(k^2-\frac{\theta''}{\theta}\bigg) u_k=0,
\eeq

the Fourier's version of \eqref{umasterinf}.

The vacuum expectation value for the operators $\hat{\Psi}(\eta,x) \hat{\Psi}(\eta,y)$  is then:

\beq\label {2pf}
\langle 0| \hat{\Psi}(\eta,x) \hat{\Psi}(\eta,y) \vac = \int \frac{dk}{k} 4G^2(\rho + P)|u_k|^2 k^3 \frac{\sin kr}{kr},
\eeq

where $r \equiv |x-y|$. Given that the definition of the power spectrum of  a gaussian random field is:

\beq
\ovl{\Psi(x,\eta) \Psi (y, \eta)} = \int  \frac{dk}{k} \mP_\Psi (k,\eta) \frac{\sin kr}{kr},
\eeq

one  can read  the power spectrum of the metric perturbations  directly from  \eqref{2pf}:

\beq\label{896}
\mP_\Psi (k,\eta) = 4G^2(\rho+P)|u_k(\eta)|^2 k^3.
\eeq

The reminder of  the calculation can be summarized as follows: given the initial conditions $v_k(\eta_i)$ and $v_k'(\eta_i)$ set by the ``Bunch Davis vacuum"
  one obtains the initial conditions  $u_k(\eta_i)$ and $u_k'(\eta_i)$ from  \eqref{857}. Given those,  one  solves \eqref{859}
and uses the resulting   function $u_k(\eta)$
 in the expression   \eqref{896} of the power spectrum.
  The  
  equation for $u_k(\eta)$ \eqref{859}
  is usually solved
  by considering
 an early situation  where  the physical scale of the mode
  is ``well inside of the Hubble radius" $k \gg aH$ (short-wavelength) and then connecting  the solution to that corresponding to the  latter regime  where
  the physical scale of the mode is ``outside of the Hubble radius" $k \ll aH$ (long-wavelength).


The  expressions for  $ u_k(\eta) $ in the short-wavelength and long-wavelength regimes are, respectively:

\beq\label{ukcortaf}
u_k(\eta) = \frac{-i}{k^{3/2}}e^{ik (\eta-\eta_i)}   \qquad \qquad \mbox{for} \qquad k^2 \gg \frac{\theta''}{\theta},
\eeq

and

\beq\label{uklargaf}
 u_k(\eta) =\alpha_k \theta(\eta) + \frac{\beta_k \theta(\eta)}{4 \pi G}  \bigg( \frac{a^2(\eta)}{\mH(\eta)} - \int^{\eta} a^2 (\tilde{\eta}) d\tilde{\eta} \bigg) \qquad \mbox{for} \quad k^2 \ll \frac{\theta''}{\theta},
\eeq

with $\alpha_k$ and $\beta_k$  given by:

\beq\label{betak}
 \alpha_k = k^{-\half} \mpl^2 \sqrt{\frac{6\epsilon}{V}} e^{ik(\eta_k-\eta_i)}(-1+i\epsilon), \qquad \beta_k = k^{-\frac{3}{2}} \frac{e^{ik(\eta_k-\eta_i)}}{\mpl^2} \sqrt{\frac{V}{6\epsilon}} (1-i(\epsilon+1)).
\eeq

The final step is to calculate the power spectrum \eqref{896}
considering  that the relevant modes ($u_k(\eta)$) are in the long-wavelength regime during the radiation-dominated epoch ($a(\eta) \propto \eta$).
 Thus  neglecting  the first term in  \eqref{uklargaf} ($\theta$ is inversely proportional to $a(\eta)$ which is  a rapidly
increasing function),  we can express  $u_k(\eta)$ as:

 \beq\label{uklargarad}
u_k(\eta) \approx \frac{2}{3}
\frac{\beta_k}{4\pi G (\rho + P)^{\half}}.
\eeq


Substituting  \eqref{uklargarad} into  \eqref{896} one obtains:

\beq\label{psrad}
\mP_\Psi (k,\eta) = 4G^2(\rho+P)|u_k(\eta)|^2 k^3 \approx \frac{V}{27 \pi^2 \mpl^4 \epsilon},
\eeq

where we have assumed   smallness of the slow-roll parameter $\epsilon \ll 1$, obtaining a scale-invariant power spectrum compatible with the flat Harrison-Zeldovich spectrum. 

As shown in the previous calculations, the fact that $\mP_\Psi$ is proportional to $V/\mpl^4\epsilon$ can be traced back to the   late time behaviour of $u_k(\eta)$,
as shown in   \eqref{uklargarad}: It
 is proportional  to $|\beta_k|^2$  which is directly proportional to $V/\mpl^4\epsilon$.

Thus we can trace the  result regarding the  amplitude of the spectrum,  to the behaviour
 $u_k(\eta)$ in   the long-wavelength approximation  during the radiation dominated epoch, considered  as the   time  where the imprint of these inhomogeneities on the CMB occurs,  and to the  matching  conditions  for this  quantity   at the  time of ``horizon crossing".

This  result is thus,  closely connected  to  the  identifications  made in the standard approach to the problem of the quantum-to-classical transition, which  we regard  as  very problematic  at the conceptual level, as we have  already pointed out.

 Next   we  will see how  our  approach, which is   completely different at the fundamental level, can  in principle change the result \eqref{psrad}, but  we  will find that the  required  characterization of   the collapse  contains  elements that  seem to make it  unphysical.


\section{Earlier Arguments}\label{Earlier}

In the  standard approach one can  trace the   problem  to  an  enormous  amplification of the  fluctuation spectrum occurring during the transition from the inflationary regime to the radiation dominated regime. In order to consider  the  ratio  by which $\Psi_k(\eta)$ is amplified 
during the transition,  one  focusses  attention on $\xi$  the  so called ``intrinsic curvature perturbation",  the  quantity that can be  regarded in this context as   defined by:

\beq\label{xi}
\xi \equiv  \frac{2}{3} \frac{\mH^{-1} \Psi' + \Psi}{1+w} + \Psi,
\eeq

where $w\equiv P/\rho$. This quantity
was first introduced in \cite{bardeen}  and has  been  extensively used \cite{kahn,lyth}.
It  can  be shown, by using the definition of $c_s^2 \equiv P'/\rho'$, the continuity equation $\rho'=-3\mH(\rho+P)$ and with the help of
 \eqref{psimaster},  that $\xi$ is, in the  long wavelength approximation and for ``adiabatic perturbations",  roughly  a ``constant  quantity", irrespective of the cosmological  regime and  the nature of the dominant kind of matter.
The   constancy of  this   quantity  during the transition from the inflationary epoch to the radiation dominated epoch,
is used  to obtain  a relation between  the values of the Newtonian potential  during the  two  relevant regimes:  $\Psi^{inf}_k(\eta)$ and $\Psi^{rad}_k(\eta)$.

\beq\label{xib}
 \xi^{inf} =  \xi^{rad} \qquad \mbox{implies} \qquad \Psi^{inf}_k \bigg[ \frac{2}{3} \bigg( \frac{1}{w_{inf} + 1} \bigg) +1\bigg] = \frac{3}{2} \Psi^{rad}_k,
\eeq

where,  in obtaining  the right hand side of   \eqref{xib} the use  of  the equation of state  $P = \rho/3$ was made,  and  the left hand side was obtained  using the equation of state $P = w_{inf} \rho$ where $w_{inf} + 1 = \phi_0'^2/ a^2 \rho$. We should note that the constancy of the  Newtonian potential (for scales larger than the Hubble radius) within the   inflationary and  radiation dominated regimes,  was   used in the form $\Psi_k'=0$ in both sides of the equation. Finally   by relying on  the   assumption of validity  of the slow-roll approximation  during inflation,  $\phi_0'^2/a^2 = \frac{2}{3} V \epsilon$,  \eqref{xib} becomes:

\beq\label{psiinfrad}
\Psi^{rad}_k = \frac{2}{3} \frac{\Psi^{inf}_k}{\epsilon}.
\eeq

We note  the difference between the power spectra obtained at different cosmological epochs:

\beq
\mP_\Psi^{inf} (k,\eta) = \frac{k^3}{4\pi^2} \epsilon^2 |\beta_k|^2,  \qquad  \mP_\Psi^{rad} (k,\eta) = \frac{k^3}{4\pi^2} \frac{4}{9} |\beta_k|^2,
\eeq

and since $|\beta_k|^2 \propto k^{-3} V/\mpl^4 \epsilon$ last equations yield:

\beq
\mP_\Psi^{inf} (k,\eta) \propto \frac{V \epsilon}{\mpl^4}, \qquad  \mP_\Psi^{rad} (k,\eta) \propto \frac{V}{\epsilon \mpl^4}.
\eeq


 {\bf One can  next  consider, the  corresponding analysis  for the collapse scheme.}   The  fundamental  difference is that in this approach,  there  is  more intrinsic  freedom in the  model that goes beyond the specification  of the  set of fields  and  the potential.   One  must  characterize  the  collapse time  and the  state after the collapse. That  suggests 
  that  a specific  scheme of the collapse,   might prevent  the Newtonian  potential $\Psi_k$ from  getting amplified during the transition from the inflationary to the radiation dominated epoch.


 We   can look at   this  possibility by focusing  again  on  the quantity  $\xi$.
  During the inflationary epoch, the expression \eqref{xi} for $\xi$ can be rewritten (by using $\delta G^0_i = 8\pi G \delta T^0_i$ in its semiclassical version along with $\rho+P = \phi_0'^2/a^2$ and $1+w = 8\pi G \phi_0'^2/3\mH^2$ ) as:

\beq
\xi_{inf} =  \Psi_{inf} + \frac{\mH}{ \phi_0'}   \bra \hat{\delta\phi} (\eta)  \ket_{\Theta},
\eeq

  where the inflaton field  $\delta\phi$  has  been replaced  by  the  expectation value  of  the  corresponding  quantum operator in the  state $|\Theta \ket$ after the collapse.  Here of course,  we are dealing  with a quantum expectation value, and  one might  worry about the  appropriateness of  using  properties  derived for the corresponding  quantity in a classical  realm.  However  we  know that  the  the Schr\"odinger or Heissenberg  evolutions imply,  for this system that the expectation  values  follow  the  same  equations of motion as the classical counterparts (Ehrenfest's Theorem). In our case  of course the collapse  scheme departs  from such  smooth evolution, but as  we must focuss on the  late time  (i.e.   the  reheating  era) behaviour  of the  quantities  and assuming that the collapse  occurs  well  within the  inflationary  stage, we  would  be   justified  in taking over,  for the   behaviour of   this $ \xi $, the conclusions  obtained  from the classical   equations of motion.

\smallskip
If the post-collapse state $|\Theta \ket$, is  such that   $\bra \hat{\delta\phi} (\eta_k^c)  \ket = 0$,
 given the constancy of $\xi$ (for modes $k\eta \ll 1$) one  infers that:

\beq\label{xiinf}
\xi_{inf} = \Psi_{inf}.
\eeq

During the  radiation dominated epoch, the expression for $\xi$ is obtained from  \eqref{xi} by considering $w=\frac{1}{3}$ and the constancy of $\Psi$ in the long-wavelength regime, therefore

\beq\label{xirad}
\xi_{rad} = \frac{3}{2} \Psi_{rad}.
\eeq

Using that $\xi$ is a conserved quantity we can obtain a relation between $\Psi_{inf}$ and $\Psi_{rad}$ within the  particular collapse scheme ( \eqref{xiinf} and \eqref{xirad}):

\beq\label{xic}
\xi^{rad} =  \xi^{inf} \qquad \mbox{implies} \qquad  \Psi^{rad}_k  = \frac{2}{3} \Psi^{inf}_k.
\eeq

Thus, this last  result indicates that in the collapse framework, the amplitude of the metric perturbations would only be ``amplified" by a factor of $\frac{2}{3}$ during the transition from the inflationary to the radiation dominated
epoch,  a  result which differs drastically from the standard inflationary approach because in that case,  the amplification of the metric perturbation was of $1/\epsilon$. In the absence  of the huge  amplification,
the  final amplitude  would seem to be simply  proportional to $V\epsilon/\mpl^4$.  We will  next see  how   a more  detailed  and careful analysis   invalidates  this  happy conclusion.


\section{The collapse scheme for the quantum fluctuations in the inflationary scenario}\label{sudarsky}


In this section we will review the formalism used in analyzing the collapse process, the full formalism and motivation is exposed in \cite{sud} and \cite{sud2}.  Here  the working assumptions are the  validity of a classical treatment for  the space-time  according to  the semi-classical  Einstein's  equations:
 $G_{ab}= 8\pi G \langle \hat{T}_{ab} \rangle$, and the   standard  quantum  field theoretical treatment of the  inflaton's  perturbations,  with an appropriate  modifications to  include the  {\it self induced  collapse of the  wave function}. The latter is  assumed to induce  a jump in the  quantum state for each mode of the  scalar field  $|0\rangle \to |\Theta \rangle $,  and  the  corresponding  changes in the
the expectation of the  energy  momentum tensor,  leading  to  the  emergence  in the  metric
perturbations\footnote{The analysis  is done  by choosing a specific  gauge, and not  in terms of the so  called ``gauge invariant combinations",  because in the approach followed  here, the  metric  and field fluctuations are  treated on a different footing. The  metric  is considered a classical  variable (taken to  be describing,  in an  effective  manner,  the deeper fundamental  degrees of freedom of  the quantum gravity theory  that  one envisions, lies  underneath),  while the  mater fields, specifically the inflaton  field perturbations are given a  standard quantum  field (in curved space-time) treatment, with the two connected trough the semiclassical  Einstein's  equations.
The choice of  gauge  implies that the time  coordinate is attached  to  some specific  slicing of  the  perturbed space-time, and thus, our identification of  the corresponding  hypersurfaces (those of constant time)  as the  ones  associated with the  occurrence of collapses,--something deemed as an actual physical  change--,  turns  what is normally  a simple  choice of  gauge  into a choice of   the distinguished  hypersurfaces,  tied to the putative physical process behind  the collapse. This  naturally leads to  tensions  with the expected   general covariance of a fundamental theory, a  problem that afflicts  all known  collapse  models, and  which in the non-gravitational   settings becomes the issue   of compatibility with  Lorentz  or Poincare invariance of the proposals.  We must acknowledge that this generic problem  of  collapse  models is indeed  and   open issue for the present approach. One  would expect that its  resolution  would be tied to the  uncovering the  actual  physics  behind  what we treat here as  the  collapse of  the  wave function (we which we  view as a merely an effective description). As has been argued in related  works, and  in following ideas originally  exposed  by R. Penrose\cite{penrose},   we hold that the physics  that  lies behind  all  this,  ties the  quantum  treatment of gravitation  with the foundational issues  afflicting quantum theory in general, and in particular those  with connection  to the ``measurement problem". }.
For more details we refer the reader to \cite{sud, sud2}.

The  analysis  here will be  based on the  same  model as that of  section II. As in the usual treatment, one splits both, metric and scalar field into a spatially homogeneous ``background" part and an inhomogeneous part ``fluctuation", i.e., the scalar field is written $\phi = \phi_0 + \dphi$, while the perturbed metric (in the longitudinal gauge) can be written as in \eqref{metric}.


The equations governing the background unperturbed Friedmann-Robertson universe and the homogeneous scalar field $\phi_0(\eta)$ are again, the scalar field equation \eqref{1000} and Friedmann's equations \eqref{friedmann1} and \eqref{friedmann2}. We will work in the same set up for the values of $\eta$ and $a(\eta)$ mentioned at the beginning of section \ref{mukhanov}.

Let us begin the quantum theory within the collapse framework by incorporating the fact that
 $\Psi = \Phi$  in Einstein's equations  \eqref{00inf1} and \eqref{0iinf1}. By setting the appropriate boundary condition,  \eqref{0iinf1} reduces to
\beq\label{0iinf2}
\Psi' + \mH\Psi= 4\pi G \phi_0' \dphi.
\eeq

Then by substituting this last expression in the left hand side of \eqref{00inf1} and noting that $4 \pi G \phi_0'^2 \Psi =4\pi G a^2 (\rho + P)\Psi =  (\mH^2- \mH')\Psi$  (where the last equality was given by  \eqref{friedmann2}) one obtains:

\beq\label{nabla2psi}
\nabla^2 \Psi + \mu \Psi = 4\pi G ( \omega \dphi + \phi_0' \dphi'),
\eeq

where $\mu \equiv \mH^2 - \mH'$ and $\omega \equiv 3 \mH \phi_0' + a^2 \partial_\phi V$, which upon use of the expression for $\partial_\phi V$ from \eqref{1000}, gives $\omega = \mH \phi_0' - \phi_0''$.
The slow-roll approximation in terms of the cosmic time is given as: $\partial^2 \phi_0 / \partial t^2 \approx 0$.
This approximation can be written in terms of the conformal time as $\phi_0''- \mH \phi_0' \approx 0$, that is, $\phi_0'' \approx \mH \phi_0'$.
Then, the equation of  motion  for the background field in the slow-roll approximation is

\beq\label{srmeq}
3\mH \phi_0' + a^2 \partial_\phi V \approx 0.
\eeq

Equation \ref{srmeq} implies that $\omega \approx 0$. Thus \eqref{nabla2psi} becomes:

\beq\label{25b}
\nabla^2 \Psi +\mu \Psi =  4 \pi G \phi_0' \dphi'.
\eeq

On the other hand, the evolution equation for the fluctuation of the field  obtained from action \eqref{actioncol} is:

\beq\label{dphimot}
\dphi''-\nabla^2\dphi + 2\mH \dphi'=0.
\eeq

 Note that  here we have neglected terms proportional to $\partial^2_{\phi \phi} V[\phi]$ evaluated on ($\phi_0$ during the inflationary period),  in accordance  with  assumed  flatness of the potential in the  standard  slow-roll approximation.

It is convenient to work with the auxiliary field $y=a \dphi$, thus  \eqref{dphimot} becomes:

\beq\label{ymot}
y''- \bigg( \nabla^2 + \frac{a''}{a} \bigg) y = 0.
\eeq

The next step involves the quantization of the field fluctuation, that is, one writes the field as $\phi=\phi_0 + \dphi$, where the background field $\phi_0$ is described in a completely ``classical"\footnote{What we mean by ``classical" is that the background field $\phi_0$ is taken as an approximated description of the quantum quantity $\bra \hat{\phi_0} \ket$.}  fashion while only the fluctuation $\dphi$ is quantized.  After the quantization $\hat{\dphi} = a^{-1} \hat{y}$ of $\dphi$ it is  recognized immediately that a quantization $\hat{y}$ of $y$ occurs automatically. The conjugated canonical momentum of $y$ is $\pi = y'-ya'/a$. In order to avoid infrared problems, the collapse picture considers a restriction of the system to a box of side L, where it is imposed periodic boundary conditions. The field and its momentum is thus:

\beq\label{campomomento}
\hat{y}(\eta,\x) = \frac{1}{L^3} \sum_\nk e^{i\nk \cdot \x} \hat{y}_k(\eta), \qquad \hat{\pi}(\eta,\x) = \frac{1}{L^3} \sum_\nk e^{i\nk \cdot \x} \hat{\pi}_k(\eta).
\eeq

The wave vectors satisfy $k_i L = 2\pi n_i$ with $i=1,2,3$ and $\hat{y}_k(\eta) \equiv y_k (\eta) \hat{a}_k + \ovl{y}_k (\eta) \hat{a}_{-k}^\dag$, and $\hat{\pi}_k(\eta) \equiv g_k (\eta) \hat{a}_k + \ovl{g}_k (\eta) \hat{a}_{-k}^\dag$. The functions $y_k(\eta)$ and $g_k(\eta)$ reflect the election of the vacuum state, the so called Bunch-Davies vacuum:

\beq\label{bunchdavies}
 y_k^{\pm}(\eta)= \frac{1}{\sqrt{2k}}\bigg(1 \pm \frac{i}{\eta k} \bigg) \exp (\pm ik\eta),   \qquad g_k^{\pm} (\eta) = \pm i \sqrt{\frac{k}{2}} \exp (\pm ik\eta).
\eeq

The vacuum state is defined by the condition $\hat{a}_k \vac = 0$ for all $k$, and is homogeneous and isotropic at all scales.

The induced collapse operates in close analogy with a ``measurement" in the quantum-mechanical sense,  but of course  without any  external apparatus or observer  that could  be  thought to  make  the measurement. That is, one  assumes that at a certain time $\eta_k^c$ (from now on we will refer to this particular time as the \emph{time of collapse}) the state of the field, which was initially the Bunch Davies  vacuum   changes  spontaneously  into another state   that could in principle be a non-symmetrical state. The proposal  is inspired by Penrose's ideas \cite{penrose,penrose2} in which gravity plays a fundamental role on the collapse of the wave-function and it does not require of observers to perform a measurement in order to induce the collapse. The collapse scheme  as employed here however,  does not propose  at this point a  concrete physical  mechanism behind it,  although   one  envisions a more profound theory  presumably derived from quantum  gravity  will eventually account for it.   At this  stage one  might consider it as  a  mere  parametrization of its characteristics. These ideas and motivations are discussed in great detail in \cite{sud} and \cite{sud2}. In this paper we will only make use of the collapse scheme to calculate the power spectrum of the metric perturbations. Since the collapse acts as a sort of ``measurement", it is convenient to decompose the field $\hat{y}_k$ and its conjugated momentum $\hat{\pi}_k$ in their real and imaginary parts which are completely hermitian $\hat{y}_k(\eta)=\hat{y}_k^R(\eta)+i\hat{y}_k^I(\eta)$ and $\hat{\pi}_k(\eta)= \hat{\pi}_k^R(\eta)+i\hat{\pi}_k^I(\eta)$ and thus qualify as reasonable observables.

\begin{equation*}
 \yk^{R,I}(\eta) = \textstyle{\frac{1}{\sqrt{2}}} \left( y_k(\eta)\ak^{R,I} + \overline{y}_k(\eta)\ak^{\dag R,I} \right), \qquad   \pk^{R,I}(\eta) = \frac{1}{\sqrt{2}} \left( g_k(\eta)\ak^{R,I} + \overline{g}_k(\eta)\ak^{\dag R,I} \right),
\end{equation*}

where $\ak^R \equiv \frac{1}{\sqrt{2}}(\ak+\hat{a}_{-k}), \qquad \ak^I \equiv \frac{-i}{\sqrt{2}}(\ak-\hat{a}_{-k})$.

The commutator of the real and imaginary annihilation and creation operators  is: $[\ak^R,\aq^{R \dag}] = L^3(\delta_{k,k'}+\delta_{k,-k'}), \quad [\ak^I,\aq^{I \dag}]= L^3(\delta_{k,k'}-\delta_{k,-k'})$.

Let $|\Xi \ket$ be any state in the Fock space of $\hat{y}$. Therefore, introducing the following quantity: $d_k^{R,I} \equiv \bra \ak^{R,I} \ket_\Xi$, the expectation values of the modes are expressible as

\beq\label{490}
\bra \yk^{R,I} \ket_\Xi = \sqrt{2}Re(y_kd_k^{R,I}), \qquad \bra \pk^{R,I} \ket_\Xi = \sqrt{2} Re(g_kd_k^{R,I}).
\eeq

For the vacuum state $\vac$ one has, as expected, $d_k^{R,I}= 0$, and thus $\bra \yk^{R,I} \ket_0=0, \quad \bra \pk^{R,I} \ket_0 = 0$. While their corresponding uncertainties are: $(\Delta \yk^{R,I})^2_0=(1/2)|y_k|^2L^3, \qquad (\Delta \pk^{R,I})_0^2 = (1/2)|g_k|^2L^3$.

Now  one  needs to further   specify the state after the collapse  which  will be  denoted by $ | \Theta \ket$. For  our purposes  all it is needed to specify is $d_k^{R,I} = \bra \Theta | \ak^{R,I} | \Theta \ket$. In the vacuum state, $\hat{y}_k$ and $\hat{\pi}_k$ are individually distributed according to a Gaussian distribution centered at 0 with spread $(\Delta \yk)^2_0$ and $(\Delta \pk)^2_0$ respectively, and those  are taken as a guidance  for considering some of the characteristics of  the state  after  the collapse.
We  refer to  any such  concrete  specification  of characteristics  as a  collapse  scheme.
Various  concrete  collapse schemes,  have been studied  before,  particularly  in  \cite{adolfo}. The point is that one can test a particular collapse scheme and compare it with the observations to discard or accept it.

 In the rest of this paper we focus on  the scheme as given by \eqref{56a} and \eqref{57} because, as will be shown next, it seems to be the  one
 with the characteristics needed to deal with the amplitude of the  spectrum.  We note that  nothing of this  sort  can even be  contemplated  in the usual  approach  where  there is  no liberty whatsoever to  make  additional assumptions  at  this stage.

\beq\label{56a}
 \bra \yk^{R,I}(\eta_k^c) \ket_\Theta = \lambda_{k,1} x_{k,1}^{R,I}\sqrt{(\Delta\pk^{R,I})^2_0}=\lambda_{k,1} x_{k,1}^{R,I}|y_k(\eta_k^c)|\sqrt{L^3/2},
 \eeq

\beq\label{57}
\bra \pk^{R,I}(\eta_k^c) \ket_\Theta = \lambda_{k,2}  x_{k,2}^{R,I}\sqrt{(\Delta\pk^{R,I})^2_0}=\lambda_{k,2} x_{k,2}^{R,I}|g_k(\eta_k^c)|\sqrt{L^3/2},
 \eeq

where $\lambda_{k,1}$ and $\lambda_{k,2}$ represent real parameters; $\eta_k^c$ represents the time of collapse for the mode $k$ and $x_{k,1}^{R,I}$, $x_{k,2}^{R,I}$ are selected randomly from within a Gaussian distribution centered at zero with spread one. Here, we must emphasize that our universe corresponds to a single realization of these random variables, and thus each of these quantities has a single specific value.

Now,  as indicated  at the  beginning of this  section one  relies on a  semi-classical description of gravitation
in interaction with quantum fields,  in  terms of the semi-classical Einstein's equation $G_{ab} = 8\pi G \bra \hat{T}_{ab} \ket$ whereas the other fields are treated in the standard quantum field theory (in curved
space-time) fashion. Putting  these elements  together for the situation at hand,  the semi-classical version of the perturbed Einstein's equations \eqref{0iinf2}, \eqref{25b} valuated at the time of collapse, after a Fourier's decomposition reduce to\footnote{One  might  want to be careful if the situation involves an extreme case of slow roll condition where $\phi_0'$ was so small that higher order terms  would  dominate. In this manuscript we  will be assuming we are not facing that situation.}:

\beq\label{0iinf3}
\Psi_k'(\eta^c_k) + \mH\Psi_k(\eta^c_k)= 4\pi G \frac{\phi_0'}{a} \bra \hat{y}_k(\eta^c_k) \ket,
\eeq

\beq\label{poisson}
-k^2 \Psi_k(\eta^c_k) +\mu \Psi_k (\eta^c_k)=  4 \pi G \frac{\phi_0'}{a} \bra \hat{\pi}_k(\eta^c_k) \ket.
\eeq

It  is easy to  see that before the collapse occurs, the expectation value on the right hand side of the latter expressions is zero, and the  space-time is  homogeneous and isotropic (at the corresponding scale).

The expressions \eqref{0iinf3} and \eqref{poisson} where obtained from Einstein's equations with components $\delta G^0_0 = 8 \pi G \delta T^0_0$ and $\delta G^0_i = 8 \pi G \delta T^0_i$. It is a known result, that these particular equations are not actual motion equations but rather constraint equations. The motion equation is the one given by $\delta G^i_j = 8 \pi \delta T^i_j$ which, after use of the constraint equation \eqref{00inf1} and during the inflationary period,  is represented in \eqref{psimasterinf1} (or equivalently by \eqref{umasterinf}).

Therefore, we can use the constraint equations \eqref{0iinf3} and \eqref{poisson}, to obtain the input data for  \eqref{psimasterinf1}, i.e.,  \eqref{0iinf3} and \eqref{poisson} allow us to specify $\Psi(\eta_k^c)$ and $\Psi'(\eta_k^c)$ which will serve as the initial conditions for  \eqref{psimasterinf1}. Thus, assuming that the time of collapse $\eta_k^c$ occurs at the very early stages of the inflationary regime, for which the modes of interest satisfy $k^2 \gg \epsilon \mH^2(\eta_k^c) = \mu  \Rightarrow |k\eta_k^c| \gg \epsilon$, then \eqref{poisson} yields the initial condition:

\beq\label{poisson2}
\Psi(\eta^c_k) =  -4 \pi G \frac{\phi_0'(\eta^c_k)}{a(\eta^c_k)k^2} \bra \hat{\pi}_k(\eta^c_k) \ket.
\eeq

Using \eqref{poisson2},  \eqref{0iinf3} yields the second initial condition:

\beq\label{0iinf4}
\Psi_k'(\eta^c_k) = 4\pi G \frac{\phi_0'(\eta^c_k)}{a(\eta^c_k)}\bigg[ \bra \hat{y}_k(\eta^c_k) \ket +  \frac{\mH_c}{k^2} \bra \hat{\pi}_k(\eta^c_k), \ket \bigg],
\eeq

where $\mH_c$ denotes $\mH$ evaluated at the time of collapse $\eta^c_k$.

With the initial conditions at hand, one proceeds to solve  \eqref{psimasterinf1} or equivalently  \eqref{umasterinf}:

\begin{equation*}
u''_k(\eta)+ \bigg(k^2 - \frac{\theta''}{\theta} \bigg) u_k(\eta) = 0.
\end{equation*}

During inflation,  the quantity $\theta$  is given by $\theta = 1/\sqrt{2 \epsilon} \mpl a$. Considering that in the slow-roll approximation $1\gg \epsilon \approx$ const.,  one finds:

\beq\label{thetabiprima}
\frac{\theta''}{\theta} = \epsilon \mH^2 = \frac{\epsilon}{(1-\epsilon)^2 \eta^2} = \frac{\epsilon + \mathcal{O}(\epsilon^2)}{\eta^2} \approx \frac{\epsilon}{\eta^2},
\eeq

where in the second equality we used the expression for $\mH$ during inflation (this was introduced at the beginning of section II). Therefore,  \eqref{umasterinf} takes the form:

\beq\label{umasterinf2}
u''_k(\eta)+ \bigg(k^2 - \frac{\epsilon}{\eta^2} \bigg) u_k(\eta) = 0.
\eeq

The general solution of \eqref{umasterinf2} is given by:

\beq\label{solumasterinf}
u_k(\eta) = C_1 \sqrt{-\eta} J_\nu (-k\eta) + C_2 \sqrt{-\eta} Y_\nu (-k\eta),
\eeq

where $\nu= \half \sqrt{1+4\epsilon}$ and $J_\nu$, $Y_\nu$ correspond to  Bessel's functions of the first and second kind respectively. Using the definition of $u\equiv \Psi/(4\pi G \sqrt{\rho  + p})$, we obtain the \emph{exact} expression for the ``Newtonian Potential" in the inflationary regime:

\beq\label{psiinfcol}
\Psi_k^{inf} (\eta) =  \frac{s}{a} \bigg( C_1 \sqrt{-\eta} J_\nu (-k\eta) + C_2 \sqrt{-\eta} Y_\nu (-k\eta) \bigg),
\eeq

where $s\equiv 4 \pi G \phi_0'$. After imposing the initial conditions \eqref{poisson2} and \eqref{0iinf4} to the general solution and using the collapse scheme as introduced in \eqref{56a} and \eqref{57}, we find $C_1$ and $C_2$.

\barr\label{c1}
 C_1 &=& \bigg(\frac{L}{k}\bigg)^{\frac{3}{2}} \frac{\pi\sqrt{k}}{4} \bigg\{ \lambda_{k,1} (x_{k,1}^R + i x_{k,1}^I)   \sqrt{|z_k|} Y_\nu (|z_k|) + \nonumber \\
 & & +  \lambda_{k,2} (x_{k,2}^R + i x_{k,2}^I)  \bigg[ \bigg( \frac{1}{1-\epsilon} - \nu - \half \bigg) \frac{Y_\nu (|z_k|)}{\sqrt{|z_k|}} + \sqrt{|z_k|} Y_{\nu+1} (|z_k|) \bigg] \bigg\},
\earr

\barr\label{c2}
 C_2 &=& - \bigg(\frac{L}{k} \bigg)^{\frac{3}{2}} \frac{\pi\sqrt{k} }{4} \bigg\{ \lambda_{k,1} (x_{k,1}^R + i x_{k,1}^I)   \sqrt{|z_k|} J_\nu (|z_k|) + \nonumber \\
 & & + \lambda_{k,2} (x_{k,2}^R + i x_{k,2}^I)  \bigg[ \bigg( \frac{1}{1-\epsilon} - \nu - \half \bigg) \frac{J_\nu (|z_k|)}{\sqrt{|z_k|}} + \sqrt{|z_k|} J_{\nu+1} (|z_k|) \bigg] \bigg\},
\earr

where $z_k \equiv k \eta^c_k$ and we used that at the time of collapse $\mH_c = -1/(1-\epsilon) \eta^c_k = k / (1-\epsilon)|z_k|$. Since we are assuming that the time of collapse occurs during the early inflationary period, then $z_k$ is in the range $-\infty < z_k \ll  k\eta_r$ (we recall that
$\eta_r \approx -10^{-22}$ Mpc, therefore $z_k < 0$).

The result \eqref{psiinfcol} represents the dynamical evolution of $\Psi_k(\eta)$ during the inflationary regime within the collapse framework. In order to obtain a predicted power spectrum and contrast it with the observations, we strictly can not use the $\Psi_k(\eta)$ as given in  \eqref{psiinfcol} because it was obtained using $a(\eta)$ in the inflationary epoch. A  realistic analysis requires  us to focus  on  $\Psi_k(\eta)$ during the radiation dominated regime. Therefore, the next step is to obtain  $\Psi_k(\eta)$ during that epoch.

 In order to do this   we must again  connect the two regimes. This  is  a point  where the analysis  necessarily deviates from  what is  usually done, as  here we have,  even  before  inflation ends, actual inhomogeneities and anisotropies in the  metric.  The so  called  ``reheating  regime", where  the  inflaton field  decays  into ordinary  particles  and photons  (including   presumably dark matter particles)  {\bf is a  complicated  and  not fully understood process, quite likely involving  huge entropy creation and other complexities.  These complications are often ignored in the literature and we will do  likewise.} However,  what seems  rather clear  is that the  metric  perturbations should  be  considered  as evolving continuously  during this  regime, and to the extent that  we ignore reheating period's  temporal extent,    the  matching of  the Newtonian potential  should  be continuous  for each mode.

  After the inflationary regime has ended, the dynamical evolution of the metric perturbation would be connected  by the fluctuations of the radiation energy density, the equations that drive the evolution are naturally, the Einstein Field Equations.

Let us recast  \eqref{psimaster}  which  describes  such  situation:

\begin{equation*}
\Psi''-c_s^2 \nabla^2 \Psi + 3 \mH (1+c_s^2)\Psi' + [(2 \mH' + \mH^2 (1+3c_s^2)] \Psi = 4 \pi Ga^2 \tau \delta S.
\end{equation*}

We note that, as showed in the beginning of section \ref{mukhanov},   \eqref{psimaster} is valid for any cosmological period. In a radiation dominated universe $P=\rho/3$, therefore $c_s^2=\frac{1}{3}$ and $\tau \delta S=0$. Given the equation of state, the scale factor can be calculated using the background equations, obtaining $a(\eta) = C_{rad}(\eta - \eta_r) + a_{r}$ where: $C^2_{rad} \equiv \frac{8}{3} \pi G \rho a^4$ is a constant; $\eta_r$ is the conformal time at which the radiation epoch starts; $a_{r}$ is the value of the scale factor at $\eta_{r}$ and $\eta_r < \eta < \eta_{eq}$ with $\eta_{eq}$ the conformal time at which the universe is populated by matter-radiation equally. According to the  comments  above,  we take  the value $\eta_r$ at which the inflationary regime ends to be the same for which the radiation dominated epoch starts,  with $a_r$ representing  value of the scale factor at that time.

  For the  epoch corresponding  to a  radiation dominated universe, the Fourier transform version of \eqref{psimaster} takes the form:

\beq\label{eqpsirad2}
\Psi_k''+ \frac{4}{\eta-\eta_r + D_{rad}} \Psi_k'  + \frac{1}{3} k^2 \Psi_k = 0,
\eeq

where $D_{rad} \equiv a_{r}/C_{rad}$. The analytical solution to \eqref{eqpsirad2} is thus:

\barr\label{psiradcol}
 \Psi_k^{rad} (\eta) &=& \frac{3}{(k\eta - \zeta_k)^2} \bigg[ C_3 \bigg( \frac{\sqrt{3}}{(k\eta-\zeta_k)} \sin \bigg( \frac{k\eta - \zeta_k}{\sqrt{3}} \bigg) - \cos \bigg( \frac{k\eta - \zeta_k}{\sqrt{3}} \bigg) \bigg) \nonumber \\
 & & + C_4 \bigg( \frac{\sqrt{3}}{(k\eta-\zeta_k)} \cos \bigg( \frac{k\eta - \zeta_k}{\sqrt{3}} \bigg) + \sin \bigg( \frac{k\eta - \zeta_k}{\sqrt{3}} \bigg) \bigg) \bigg],
\earr

where $\zeta_k \equiv k \eta_r - k D_{rad}$. We see that $\Psi_k(\eta)$ as given by \eqref{psiradcol} contains the denominator $(k\eta-\zeta_k)^{-1}$, and one   might  worry  about  it becoming   singular,  but  noting that it arises  essentially from  $C_{rad}/ka(\eta)$,  the fact that  $a(\eta) \neq 0$, and  moreover is an increasing function  there  is no possibility of  a ``blow up" type of  behaviour. The constants $C_3$ and $C_4$ will be obtained by approximating the continuous change of the equation of state by a sharp jump. Therefore, the matching conditions for $\Psi$ and $\Psi'$ will be derived by rewriting the motion equation for $u$ (\eqref{umaster}, assuming adiabatic perturbations)  in the following form:

\beq\label{ukinf}
\bigg[\theta^2 \bigg(\frac{u_k}{\theta}\bigg)' \bigg]' =- k^2 c_s^2 \theta {u_k}.
\eeq

The quantity $u/\theta$ is continuous because the scale factor $a$ and the energy density $\rho$ are both continuous, the Newtonian Potential $\Psi$ does not jump during the transition. Integrating \eqref{ukinf} from $\eta_r-\delta$ to $\eta_r+\delta$, where $\delta$ is positive real number and very small in absolute terms, we obtain:

\beq
 \bigg[\theta^2_{rad} \bigg(\frac{u_{ k, rad}}{\theta_{rad}}\bigg)' \bigg] \bigg|_{\eta_r+\delta} - \bigg[\theta^2_{inf} \bigg(\frac{u_{k, inf}}{\theta_{inf}}\bigg)' \bigg] \bigg|_{\eta_r-\delta}=-k^2 \int_{\eta_r-\delta}^{\eta_r+\delta} c_s^2 \theta {u_k} d\eta.
\eeq

Assuming $\delta \rightarrow 0$, then the matching conditions given for $\Psi_k$ are:

\beq\label{matching}
 \theta^2_{inf} \bigg(\frac{u_{k, inf}}{\theta_{inf}}\bigg)'\bigg|_{\eta=\eta_r} = \theta^2_{rad} \bigg(\frac{u_{k, rad}}{\theta_{rad}}\bigg)'\bigg|_{\eta=\eta_r},  \qquad \Psi^{inf}_k(\eta_r) = \Psi^{rad}_k (\eta_r).
\eeq

We note that these conditions are equivalent to the Deruelle-Mukhanov conditions  obtained in \cite{deruelle}. From  these conditions, one can easily find the value of  the constants $C_3$ and $C_4$:

\barr\label{c3}
C_3 = &- \bigg[ \bigg(\frac{D_k^2}{3}-3\bigg) \cos \bigg(\frac{D_k}{\sqrt{3}} \bigg) - \sqrt{3} D_k \sin \bigg( \frac{D_k}{\sqrt{3}} \bigg) \bigg] A(k\eta_r,z_k) \nonumber \\
&+ \bigg[ \frac{D_k}{\sqrt{3}} \cos \bigg( \frac{D_k}{\sqrt{3}} \bigg) + \frac{D_k^2}{3} \sin \bigg( \frac{D_k}{\sqrt{3}} \bigg) \bigg] \frac{\sqrt{3}}{k} B(k\eta_r,z_k),
\earr

\barr\label{c4}
C_4 = &\bigg[ \bigg(\frac{D_k^2}{3}-3\bigg) \sin \bigg(\frac{D_k}{\sqrt{3}} \bigg) + \sqrt{3} D_k \cos \bigg( \frac{D_k}{\sqrt{3}} \bigg) \bigg] A(k\eta_r,z_k) \nonumber \\
 &- \bigg[ \frac{D_k}{\sqrt{3}} \sin \bigg( \frac{D_k}{\sqrt{3}} \bigg) - \frac{D_k^2}{3} \cos \bigg( \frac{D_k}{\sqrt{3}} \bigg) \bigg] \frac{\sqrt{3}}{k} B(k\eta_r,z_k),
\earr

where

\barr\label{a}
 A(k\eta_r,z_k) &\equiv& \frac{-s\pi}{4a} \bigg(\frac{L}{k} \bigg)^{\frac{3}{2}} \bigg\{ \lambda_{k,1} (x_{k,1}^R + i x_{k,1}^I) \bigg( 1+ \frac{1}{z_k^2} \bigg)^{\half} \sqrt{|z_k|}  \bigg[  J_\nu (|z_k|) Y_\nu (-k\eta_r)- \nonumber \\
 & & -  Y_\nu (|z_k|) J_\nu (-k\eta_r) \bigg] +  \lambda_{k,2} (x_{k,2}^R + i x_{k,2}^I)  \bigg[-\bigg( \sqrt{|z_k|} Y_{\nu+1} (|z_k|)  \nonumber \\
 & & + \bigg( \frac{1}{1-\epsilon} - \nu - \half \bigg) \frac{Y_\nu (|z_k|)}{\sqrt{|z_k|}}
 \bigg)  J_\nu (-k\eta_r) + \bigg( \bigg( \frac{1}{1-\epsilon} - \nu - \half \bigg) \frac{J_\nu (|z_k|)}{\sqrt{|z_k|}} + \nonumber \\
 & & + \sqrt{|z_k|} J_{\nu+1} (|z_k|) \bigg) Y_\nu (-k\eta_r) \bigg] \bigg\} \sqrt{-k\eta_r},
\earr

\beq
 B(k\eta_r,z_k) \equiv \frac{2}{\epsilon} \bigg[ -k \frac{\partial A(-k\eta_r,z_k)}{\partial (-k \eta_r)} + \mH(\eta_r) A(-k\eta_r,z_k) \bigg],
\eeq

and $D_k \equiv k D_{rad}$. The quantity $A(k\eta_r,z_k)$ can be approximated, by considering that if $\epsilon \ll 1$ then $\nu \approx \half$, thus we have

\barr\label{a2}
 A(k\eta_r,z_k) &\approx& \frac{-s}{2a} \bigg(\frac{L}{k} \bigg)^{\frac{3}{2}} \bigg[  \lambda_{k,1} (x_{k,1}^R + i x_{k,1}^I) \bigg( 1 + \frac{1}{z_k^2} \bigg)^\half \sin \Delta_r +  \nonumber\\
 & & + \lambda_{k,2} (x_{k,2}^R + i x_{k,2}^I)  \bigg( \cos \Delta_r  + \frac{\sin \Delta_r }{z_k} \bigg) \bigg],
\earr

\barr\label{b2}
 B(k\eta_r,z_k) &\approx& \frac{s}{2a} \bigg(\frac{L}{k} \bigg)^{\frac{3}{2}} \frac{2k}{\epsilon} \bigg\{  \lambda_{k,1} (x_{k,1}^R + i x_{k,1}^I) \bigg( 1 + \frac{1}{z_k^2} \bigg)^\half \bigg( \cos \Delta_r  - \frac{\sin \Delta_r }{k\eta_r} \bigg) \nonumber \\
 & & +   \lambda_{k,2} (x_{k,2}^R + i x_{k,2}^I)  \bigg[ \cos \Delta_r  \bigg(\frac{1}{k\eta_r}- \frac{1}{z_k} \bigg)  + \sin \Delta_r  \bigg( \frac{1}{k\eta_r z_k} + 1 \bigg) \bigg] \bigg\},
\earr



where $\Delta_r \equiv k\eta_r - z_k$. Despite the apparent complexity for the constants $C_3$ and $C_4$, we note  that for the scales of interest $10^{-3} \mbox{ Mpc}^{-1} < k < 1 \mbox{ Mpc}^{-1}$  and $D_{rad} = a_{r}/C_{rad} \approx 1.5 \times 10^{-22} \mbox{ Mpc}$, we have $ D_k \in [10^{-25},10^{-22}]$. Therefore,  the approximated expressions for \eqref{c3} and \eqref{c4} up to first order in $D_k$ are:

\beq\label{c3a}
C_3 \approx 3 A(k\eta_r,z_k) + \frac{D_k}{k} B(k\eta_r,z_k),
\eeq

\beq\label{c4a}
C_4 \approx 0.
\eeq

We should note that the approximation given by the expressions above would correspond, in the standard treatment, to neglecting the ``decaying mode".  Here  we  can see clearly   how this  comes about  as a result  of the matching conditions and the   range of values of   the relevant  quantities.

With this  results  we  can   now   write an  approximate expression for   the  Newtonian Potential, during the radiation dominated epoch,  as:

\barr\label{psiradap}
 \Psi^{rad}_k(\eta) &\approx& \bigg[ 3 A(k\eta_r,z_k) + \frac{D_k}{k} B(k\eta_r,z_k) \bigg]   \frac{3}{(k\eta - \zeta_k)^2} \bigg[ \frac{\sqrt{3}}{(k\eta-\zeta_k)} \sin \bigg( \frac{k\eta - \zeta_k}{\sqrt{3}} \bigg) - \nonumber \\
 & & - \cos \bigg( \frac{k\eta - \zeta_k}{\sqrt{3}} \bigg) \bigg],
\earr

where the constants $A(k\eta_r,z_k)$ and $B(k\eta_r,z_k)$ are given by \eqref{a2} and \eqref{b2} respectively.

Next we turn to the observational quantities. The quantity that is measured is $\Delta T(\theta, \varphi)/T $ which is a function of the coordinates on the celestial two-sphere and is expressed as $\sum_{lm} \alpha_{lm} Y_{lm} (\theta,\varphi)$. The angular variations of the temperature are then identified with the corresponding variations in the ``Newtonian Potential" $\Psi$, by the understanding that they are the result of gravitational red-shift in the CMB photon frequency $\nu$ so $\delta T / T = \delta \nu / \nu = \delta (\sqrt{g_{00}}) / \sqrt{g_{00}} \approx \Psi$

The quantity that is presented as the result of observations is $OB_l = l(l+1)(2l+1)^{-1} \sum_m |\alpha_{lm}^{obs}|^2$. The observations indicate that (ignoring the acoustic oscillations, which is anyway an aspect that is not being considered in this work) the quantity $OB_l$ is essentially independent of $l$ and this is interpreted as a reflection of the ``scale invariance" of the primordial spectrum of the fluctuations.

The quantity  of observational interest is the ``Newtonian potential" on the surface of last scattering: $\Psi(\eta_D,\x_D)$, from where one extracts

\beq
\alpha_{lm} = \int d^2 \Omega \quad \Psi(\eta_D, \x_D) Y_{lm}^\star (\theta,\varphi),
\eeq

with $\x_D = R_D(\sin \theta \sin \varphi, \sin \theta \cos \varphi, \cos \theta)$ and $R_D$ represents the radius of the surface of last scattering. In order to evaluate the expected value for the quantity of interest,  we  will first use the Fourier's decomposition of the metric's perturbation:

\beq\label{69}
\Psi(\eta,\x) = \sum_\nk \frac{1} {L^3} \Psi_k(\eta) e^{i\nk \cdot \x}.
\eeq

 With the  expressions at hand, and  after some algebra, one obtains an expression for $a_{lm}$:

\beq\label{ALFA}
 \alpha_{lm} =\sum_\nk \frac{-s}{2 a_{r}(L k)^{\frac{3}{2}}}   \bigg[ 3 F(k\eta_r,z_k) - \frac{2D_k}{\epsilon} G(k\eta_r,z_k) \bigg] E(k\eta_D,k\eta_r) 4\pi i^l j_l (k R_D) Y_{lm}^{\star} (\hat{k}),
\eeq

where

\barr\label{f}
 F(k\eta_r,z_k) &\equiv& \lambda_{k,1} (x_{k,1}^R + i x_{k,1}^I) \bigg( 1 + \frac{1}{z_k^2} \bigg)^\half \sin \Delta_r + \nonumber \\
 & & +  \lambda_{k,2} (x_{k,2}^R + i x_{k,2}^I)  \bigg( \cos \Delta_r  + \frac{\sin \Delta_r }{z_k} \bigg),
\earr

\barr\label{g}
 G(k\eta_r,z_k) &\equiv& \lambda_{k,1} (x_{k,1}^R + i x_{k,1}^I) \bigg( 1 + \frac{1}{z_k^2} \bigg)^\half \bigg( \cos \Delta_r  - \frac{\sin \Delta_r }{k\eta_r} \bigg) + \nonumber \\
 & & +  \lambda_{k,2} (x_{k,2}^R + i x_{k,2}^I)  \bigg[ \cos \Delta_r  \bigg(\frac{1}{k\eta_r}- \frac{1}{z_k} \bigg)  + \sin \Delta_r  \bigg( \frac{1}{k\eta_r z_k} + 1 \bigg) \bigg],
\earr

\beq\label{ckradap}
 E(k\eta,k\eta_r) \equiv  \frac{3}{(k\eta - \zeta_k)^2} \bigg[ \frac{\sqrt{3}}{(k\eta-\zeta_k)} \sin \bigg( \frac{k\eta - \zeta_k}{\sqrt{3}} \bigg) - \cos \bigg( \frac{k\eta - \zeta_k}{\sqrt{3}} \bigg) \bigg],
\eeq

$j_l(x)$ is the spherical Bessel function of the first kind, and where $\hat{k}$ indicates the direction of the vector $\nk$.  
The above quantity should be evaluated at the conformal time of decoupling $\eta_D$ which lies  in the matter dominated epoch.
Nevertheless,   we have used the expression for $\Psi_k(\eta)$ in the radiation dominated era  \eqref{psiradcol}, extending if one  wants  the range of validity for  \eqref{ALFA} which is from $\eta_r$ to $\eta_{eq} < \eta_D$.  The changes  during the  brief  period from  the start of  ``matter domination" to   ``decoupling" (where the scale factor  changes  only by a factor of 10, i.e. $a(\eta_D)/a(\eta_{eq}) \approx 10$), are   naturally considered to  be  irrelevant for the  issues concerning us here, and thus the approximated value for $\alpha_{lm}$ obtained using \eqref{psiradcol}  should be a  very  good approximation for the exact value of $\alpha_{lm}$.

 We note here  that  the slow roll  parameter $\epsilon$ has entered  in   the   denominator  of one of the terms  in  \eqref{ALFA}, and   as we  will see  latter  one can trace  the  unwelcome  amplification of  the   overall  scale of the  fluctuation spectrum  precisely to  this factor. The possibility of   avoiding this problem in the  collapse  scheme,  which has  no counterpart in the ordinary treatments,   arises precisely from the liberty to  select the  details of the collapse  so as to ensure the vanishing of  the coefficient of the  $1/\epsilon$  in  \eqref{ALFA}.  We  will discuss this possibility  shortly.

 One should note in passing that it is in  \eqref{ALFA} that   one can find the justification for the  reliance on statistical considerations, despite the fact that we are dealing with a single universe (even if we assume that many exist the fact is that we have empirical access only to one). The quantity we are interested on,  $\alpha_{lm}$   is,  as shown in  \eqref{ALFA}, the result of the combined contributions of  the collapse of  the wave functions of an ensemble of harmonic oscillators, (one for  each one of the  quantum field  modes $\vec k$), with each one contributing with a complex number to the sum, leading to what is in effect a 2-dimensional random walk whose total displacement corresponds to the  quantity of actual  observational interest.  It is  clear that, as in the case of any random walk,  such  quantity can not  be evaluated and  the only thing  that can  be done  is to evaluate the most likely value for such total displacement,  with the expectation that the observed  quantity  will be   close to  that value.

The expected magnitude of the  quantity
$\alpha_{lm}$, after taking the continuum limit ($L \rightarrow \infty$) is (See \cite{sud} for  details):

\beq
 |\alpha_{lm}|^2_{M.L.} = \frac{s^2}{2 \pi a_{r}^2} \int \frac{d^3k}{k^3} H(k\eta_r,z_k) E^2(k\eta_D,k\eta_r) j_l^2 (k R_D) |Y_{lm} (\hat{k})|^2,
\eeq

where

\barr\label{h}
 H(k\eta_r,z_k) &\equiv& 2\lambda_{k,1}^2 \bigg( 1+\frac{1}{z_k^2} \bigg) \bigg[ 3 \sin \Delta_r - \frac{2 D_k}{\epsilon} \bigg( \cos \Delta_r - \frac{\sin \Delta_r}{k\eta_r} \bigg) \bigg]^2 + \nonumber \\
 & & + 2 \lambda_{k,2}^2 \bigg\{ 3 \bigg( \cos \Delta_r + \frac{\sin \Delta_r}{z_k} \bigg) - \frac{2D_k}{\epsilon} \bigg[ \cos \Delta_r \bigg (\frac{1}{k\eta_r} - \frac{1}{z_k} \bigg) + \nonumber \\
 & & + \sin \Delta_r \bigg( \frac{1}{k\eta_r z_k} +1 \bigg) \bigg] \bigg\}^2.
\earr

The expected  value for the observed quantity $OB_l = l(l+1)(2l+1)^{-1} \sum_m |\alpha_{lm}^{obs}|^2$ is thus:

\beq
OB_l = l(l+1)   \frac{s^2}{\epsilon^2 \pi  a_{r}^2} \int  \frac{dk}{k} H(k\eta_r,z_k)  E^2(k\eta_D,k\eta_r) j_l^2 (k R_D).
\eeq

The quantity $OB_l$ is related with the amplitude of the Newtonian potential, that is, one can extract
an ``equivalent power spectrum" for the metric perturbations:

\beq\label{pscolrad3}
\mP_\Psi^{col} (k,\eta) = \frac{s^2}{8 \pi^2 a_{r}^2} H(k\eta_r,z_k) E^2(k\eta_D,k\eta_r).
\eeq

We note that if $\mP_\Psi^{col} (k,\eta)$ is independent of $k$, then the quantity $OB_l$ is independent of $l$ in correspondence with the observations.


It is easy to show, with the help of $a(\eta)$ during the radiation dominated epoch,  that the quantity $(k\eta - \zeta_k)/\sqrt{3}$ which appears in \eqref{ckradap}, is in fact $k/\sqrt{3}\mH$. As discussed previously this quantity should be evaluated at the time of decoupling $\eta_D$. However for the modes of interest we have the following condition $k/\mH \ll 1$. This corresponds to focuss on the   so called ``scales larger than the Hubble radius". That is,   we  must consider  \eqref{ckradap}, for scales $k \ll aH$, whereby we  find  that to a good  approximation:

\beq\label{ckradap2}
E(k\eta,k\eta_r) \approx \frac{1}{3}.
\eeq

Considering \eqref{ckradap2} and recalling the definition of $s = 4\pi G \phi_0'$; the  equation  of motion for the field in the slow-roll approximation $\phi_0' = -\partial_\phi V a^3/3a'$;  Friedmann's equation in the slow-roll regime $3a'^2=8 \pi G a^4 V(\phi_0)$; the  definition of the slow-roll parameter $\epsilon = \frac{1}{2}M_{pl}^2 (\partial_\phi V/V)^2$ and the definition of the reduced Planck mass $M_{pl}^2=1/(8\pi G)$, then the overall amplitude  of the predicted power spectrum is:

\beq\label{pscolrad3}
\mP_\Psi^{col} (k,\eta) = \frac{1}{432 \pi^2} \frac{V \epsilon}{\mpl^4} H(k\eta_r,z_k).
\eeq

The quantity $H(k\eta_r,z_k)$ depends on the parameters characterizing the collapse, that is, it depends on $\lambda_{k,1}$, $\lambda_{k,2}$, and $z_k$.   We  should note that the liberty to choose   those value  corresponds to   a characterization of  some of the details  of the  mechanism of collapse, and that  no  analogous  freedoms can be identified  when  one ignores the problems  we  had mentioned  at the  beginning,   and  which motivates the proposals to  modify  the inflationary paradigm  with the
 ``collapse  of the  wave  function" hypothesis.     As  we  indicated  one  can now  assume that details of  collapse  are such  that the  undesired  terms  disappear.  For instance, by
adjusting the parameters of the collapse to be $\lambda_{k,1}=0$, $\lambda_{k,2}=1$ and the time of collapse to satisfy the following equation:

\beq\label{zk}
\cos \Delta_r \bigg (\frac{1}{k\eta_r} - \frac{1}{z_k} \bigg) + \sin \Delta_r \bigg( \frac{1}{k\eta_r z_k} +1 \bigg) = 0,
\eeq

the quantity $H(k\eta_r,z_k)$ will turn out to be just:

\beq\label{h2}
H(k\eta_r,z_k) = 18 \bigg( \cos \Delta_r + \frac{\sin \Delta_r}{z_k} \bigg)^2,
\eeq
   which  does not contain the   bothersome  terms proportional to $1/\epsilon$.
   The overall amplitude of the metric perturbations will, thus,  have the following form:

\beq\label{pscolrad4}
\mP_\Psi^{\star col} (k,\eta) = \frac{1}{24 \pi^2} \frac{V \epsilon}{\mpl^4} \bigg( \cos (k\eta_r-z_k) + \frac{\sin (k\eta_r-z_k)}{z_k} \bigg)^2,
\eeq

where the $\star$ over the $\mP_\Psi$ denotes that it is a very specific model for  the collapse which leads to  the result \eqref{pscolrad4}. The amplitude given by \eqref{pscolrad4} has the desired  feature of having the factor  $\epsilon$  in the numerator rather than in the denominator, in contrast  with  the standard inflationary results. Therefore, as the  slow  roll parameter   $\epsilon $ takes  smaller values,   the  amplitude of the fluctuation spectrum  becomes   smaller,  the complete  opposite  of what  happens  in the usual approach.  This  would be  a natural resolution of the  fine tuning problem affecting  most of  the inflationary models,  if  there was a natural way to  explain  the particular values  characterizing this collapse  scheme. In principle, and given the fact that  we  certainly do not  know the  physics  behind  the collapse (as indicated  it  seems likely that  it might be connected with aspects of quantum  gravity  as  suggested  by R. Penrose), one  would   not  concern   oneself  with this issue  at this  time. However there is a very problematic aspect of the  choice   of  parameters we  made, that seems to be shared by all the other possible choices  compatible with the   desirable behavior,  namely  that  it depends on the value of $z_k$, a quantity that  is determined  by  the   stage at which the reheating occurs, and  which is   presumed to be to the  future of the  time at  which the collapse should occur.  In other  words there is a teleological aspect in the  choice of the  collapse parameters  and it is very   hard  to see  how  could  it  be part  of any   sort  of reasonable physical process, tied or not,  with quantum gravity.

Regarding the ``scale invariance" of the spectrum, we note that if $|z_k| \gg |k\eta_r|$, which is not a strong assumption as we have  $|\eta_r| \approx 1.157 \times 10^{-22}$ Mpc, which implies $10^{-25} < |k\eta_r| < 10^{-22}$. Then, the scale dependance of the predicted spectrum will be contained only in $z_k$, leading to the conclusion that there is  one simple  way to  obtain a  flat spectrum leading to  a  matching  with the  observed shape (but  as  discussed in \cite{adolfo} this is  not  the only option),  and we  will focus on that  case here  for simplicity, and because  we  are  concerned   here with the  overall scale of the fluctuation spectrum   rather than its  shape.  The point is that  by  assuming that the  time of collapse of the different modes should depend on the mode's frequency according to $\eta_k^c=z/k$, so that $z_k$ is  independent of $k$ (see  \cite{sud}  for a   possible physical   explanation  behind   of such pattern),  we  obtain  a form of the  predicted spectrum that is in agreement with the so called scale invariant spectrum obtained in ordinary treatments and in the observational studies.

  Going  back to  the   overall scale  of the  fluctuation spectrum  we  note that for a generic collapse scheme, the term in \eqref{h} which contains the factor $1/\epsilon$ becomes dominant and the quantity $H(k\eta_r,z_k)$ can be approximated by

\barr\label{hgen}
 H(k\eta_r,z_k) &\approx& \frac{8 D_k^2}{\epsilon^2} \bigg\{ \lambda_{k,1}^2 \bigg( 1+\frac{1}{z_k^2} \bigg) \bigg( \cos \Delta_r - \frac{\sin \Delta_r}{k\eta_r} \bigg)^2 + \nonumber \\
 & &+ \lambda_{k,2}^2 \bigg[ \cos \Delta_r \bigg (\frac{1}{k\eta_r} - \frac{1}{z_k} \bigg) + \sin \Delta_r \bigg( \frac{1}{k\eta_r z_k} +1 \bigg) \bigg]^2 \bigg\}.
\earr

Inserting this last expression into \eqref{pscolrad3}, we obtain the power spectrum of the perturbations for a generic collapse:

\barr\label{pscolgen}
 \mP_\Psi^{col} (k,\eta) &=& \frac{1}{54\pi^2} \frac{V}{\epsilon \mpl^4} \bigg\{ D_k^2 \lambda_{k,1}^2 \bigg( 1+\frac{1}{z_k^2} \bigg) \bigg( \cos \Delta_r - \frac{\sin \Delta_r}{k\eta_r} \bigg)^2 + \nonumber \\
 & & + D_k^2 \lambda_{k,2}^2 \bigg[ \cos \Delta_r \bigg (\frac{1}{k\eta_r} - \frac{1}{z_k} \bigg) + \sin \Delta_r \bigg( \frac{1}{k\eta_r z_k} +1 \bigg) \bigg]^2 \bigg\},
\earr

showing that in general  the overall amplitude  for the power  spectrum   would   be proportional to $V/\epsilon \mpl^4$ as in the standard approach.






\section{Conclusions}\label{discusion}

In the standard inflationary approach, one  finds one  must un-naturally  constraint in the energy scale of $V$ because of the  fact  that the prediction  for the  amplitude  of the primordial  fluctuation  spectrum  is $\mP_\Psi^{std} \sim V/\epsilon M_{pl}^4$. This is considered as a fine tuning  problem as it  indicates that in  the  model  one  must decrease $V$  as $\epsilon$  decreases.  That is, as  one   adjust $\epsilon$ to  get a flatter spectrum  one  must  adjust the potential to prevent the  overall scale of the fluctuations   from becoming too large.
 On the other  hand the  standard approach  suffers  from some  serious conceptual shortcomings  as discussed  for instance in \cite{sud}.   We  thus   saw  as  quite hopeful  the early indications that the  modified approached  proposed to  deal  with the conceptual  shortcomings  would  naturally resolve  the  more technical,  fine tuning problem.

We have seen  here  that indeed, in  the ``collapse picture" the  prediction for the amplitude of the power spectrum can  be  $\mP_\Psi^{\star col} \sim \epsilon V/ M_{pl}^4$  if we  choose   a very particular characterization  of the collapse's parameters. For a generic collapse, however  we  do obtain the same unwelcome result as in the standard approach.    Furthermore,  we found that the   choice of the parameters that lead to the  un-amplified  spectrum,  seems to  involve an  undesirable teleological aspect.   This  is  as  far as  we can see  now  a serious blow for our early hopes in this particular regard. However, the  fact  that   the new approach involves  different aspects   as part of explanation of the birth of the  primordial inhomogeneities   might still  lead to  unexpected   possible  approaches to deal with this and  other   problematic aspects  of the inflationary models.

  It is worthwhile  noting that the collapse model was not originally conceived to  deal  with the problems   such as that of the amplitude of the spectrum,
but to deal with the conceptual  issues affecting  the standard explanation of the  origin of cosmic structure. In this sense    we  should stress that  this  work  not only  shows that  the  ideas tied to the collapse  scheme  are not mere  philosophical  in nature,  but  are susceptible to  standard  theoretical  analysis, and   indeed   the fact  that the collapse model offered, a  technical possibility  of  adjusting the parameters describing  the  collapse,  in  such a way  to eliminate the  fine tuning problem,    and despite the fact that  such  solution seems  rather unconvincing (at least in the absence of a causal mechanism that could enforce the condition of \eqref{zk}),  can be considered  as illustrative of the potential of  new approach   in dealing   with  more  specific  and  less conceptual aspects of
inflationary cosmology.

We end our discussion by noting that the inflationary scenario provides an important source of actual observational data -perhaps the only one- about the gravity/quantum interface, and if, as we  believe, these  can be  connected  to  some  aspect of quantum gravity,
the careful  study of said   issues might  at long last offer  potential clues  on a subject  generally  considered  as empirically unreachable.

\section*{Acknowledgments}
We gratefully acknowledge very useful discussions with Adolfo De Unánue, Alberto Diez-Tejedor and Roy Maartens.
This work was supported in part  by  the DGAPA-UNAM grant IN119808-3, and  by  the  Conacyt grant No 101712.


\section*{References}
\begin{thebibliography}{99}




\bibitem[1]{snova}  Perlmutter S \textit{et al} 1997 \textit{Astrophys. J.} \textbf{483} 565 \\
                    Riess A G \textit{et al} 1998 \textit{Astron. J.} \textbf{116} 1009 \\
                    Wood-Vasey W M \textit{et al} 2007 \textit{Astrophys. J.} \textbf{666} 694 (arXiv:astro-ph/0701041)




\bibitem[2]{lss} York D G \textit{et al} 2000 \textit{Astron. J.} \textbf{120 } 1579 \\
                 Stoughton C \textit{et al} 2002  \textit{Astron. J.} \textbf{123} 485 \\
                 Abazajian K \textit{et al} 2003 \textit{Astron. J.} \textbf{126} 2081 







\bibitem[3]{wmap} Spergel D N \textit{et al} 2007 \emph{Astrophys. J. Suppl.} \textbf{170} 377 (arXiv:astro-ph/0603449) 



\bibitem[4]{guth}  Guth A H 1981 \textit{Phys. Rev.} D \textbf{23} 347 

\bibitem[5]{guth2} Guth A H and  Pi S Y 1982 \textit{Phys. Rev. Lett.} \textbf{49} 1110 

\bibitem[6]{muk} Mukhanov V F,  Feldman H A  and  Brandenberger R H 1992 \textit{Phys. Rept.} \textbf{215}  203 


\bibitem[7]{sud} Perez A, Sahlmann H and Sudarsky D 2006 \textit{Class. Quantum Grav.} \textbf{23} 2317  (arXiv: gr-qc/0508100) 





\bibitem[8]{sud2} Sudarsky D 2007 \textit{J. Phys.: Conf. Ser.} \textbf{68} 012029 (arXiv:gr-qc/0612005) \\
                  Sudarsky D 2007 \textit{J. Phys.: Conf. Ser.} \textbf{66} 012037 \\
                  Sudarsky D 2007 \textit{J. Phys.: Conf. Ser.} \textbf{67} 012054  (arXiv:gr-qc/0701071) \\
                  Sudarsky D 2007  The seeds of cosmic structure as a door to quantum gravity phenomena PoS(QG-Ph)038, arXiv:0712.2795







\bibitem[9]{kiefer}  Kiefer C and  Polarski D 2009 \textit{Adv. Sci. Lett.} \textbf{2} 164 (arXiv:0810.0087v2)  

\bibitem[10]{shortcomings}  Sudarsky D 2009 Shortcomings in the Understanding of Why Cosmological Perturbations Look Classical, arXiv:0906.0315v1 


\bibitem[11]{weinberg} Weinberg S 2008 \textit{Cosmology} (New York: Oxford University Press) p 476 

\bibitem[12]{mukbook}  Mukhanov V 2005 \textit{Physical Foundations of Cosmology} (New York: Cambridge University Press) 



\bibitem[13]{penrose}  Penrose R 1989 \textit{The Emperor's New Mind} (New York: Oxford University Press) 

\bibitem[14]{penrose2} Penrose R 1996 \textit{Gen. Rel. Grav.} \textbf{28} 581 


\bibitem[15]{bruna} Bruna L and Girbau J 1999 \textit{J. Math. Phys.} \textbf{40} 5131

\bibitem[16]{losic} Losic B and Unruh W G 2008 \textit{Phys. Rev. Lett.} \textbf{101} 111101 (arXiv:0804.4296)


\bibitem[17]{malik} Christopherson A J and  Malik K A 2009  \textit{Physics Letters} B \textbf{675} 159 



\bibitem[18]{davies}  Birrel N D and Davies P C W 1994 \textit{Quantum fields in curved space} (Cambridge: Cambridge University
Press) 



\bibitem[19]{bardeen}  Bardeen J, Steinhardt P and Turner M 1983 \textit{Phys. Rev.} D \textbf{28} 679 

\bibitem[20]{kahn} Brandenberger R and Kahn R 1984 \textit{Phys. Rev.} D \textbf{29} 2175 

\bibitem[21]{lyth} Lyth D 1985 \textit{Phys. Rev.} D \textbf{31} 1792 


\bibitem[22]{adolfo}   De Unanue A and  Sudarsky D  2008 \textit{Phys. Rev.} D \textbf{78} 043510  (arXiv:0801.4702v2) 

\bibitem[23]{deruelle} Deruelle N and  Mukhanov V F 1995 	\textit{Phys. Rev.}  D \textbf{52} 5549 






\end{thebibliography}
\end{document}